\documentclass[aps,prc,onecolumn,superscriptaddress]{revtex4}
\usepackage[latin1]{inputenc}
\usepackage{bm}
\usepackage{epsfig}
\usepackage{amsthm}
\usepackage{amsfonts}
\usepackage{float}
\usepackage{amsmath,amssymb}
\usepackage{color}
\usepackage{xcolor}
\usepackage{slashed}
\usepackage{relsize}
\usepackage[toc,page]{appendix} 
\usepackage{wrapfig}
\usepackage{graphicx}
\usepackage{dcolumn}
\usepackage{bm}
\newcommand{\be}{\begin{equation}} 
\newcommand{\ee}{\end{equation}} 
\newcommand{\bea}{\begin{eqnarray}} 
\newcommand{\eea}{\end{eqnarray}} 
\newcommand{\nn}{\nonumber} 
\newcommand{\mintedim}[2]{{\int\kern-0.50em\mbox{{\small$\mathop{\frac{\mbox{{\small${\rm d^{#2}}\vect{#1}$}}}{\mbox{{\small$(2\pi)^{#2}$}}}}$}}\ }} 
\newcommand{\inteonedim}[1]{{\int_0^\infty\kern-1em\mbox{{\small${\rm d}{#1}$}}}} 
\newcommand{\vect}[1]{\bm{#1}} 

\begin{document}
\title{From Harmonic Oscillators to Multiplicity Distributions in Particle Production Processes}

\author{Trambak Bhattacharyya}
\email{trambak.bhattacharyya@ujk.edu.pl}
\affiliation{Institute of Physics, Jan Kochanowski University, Kielce 25-406, Poland}

\author{Maciej Rybczy\'{n}ski}
\email{maciej.rybczynski@ujk.edu.pl}
\affiliation{Institute of Physics, Jan Kochanowski University, Kielce 25-406, Poland}

\author{Zbigniew W{\l}odarczyk}
\email{zbigniew.wlodarczyk@ujk.edu.pl}
\affiliation{Institute of Physics, Jan Kochanowski University, Kielce 25-406, Poland}

\begin{abstract} 
We propose an oscillator framework of studying multi-particle production processes using the information theory approach. For a system of degenerate oscillators, we derive the negative binomial distribution (NBD) and its nonadditive generalization represented as gNBD. We observe that gNBD derived from nonadditive statistics describes data in higher multiplicity region, thereby providing an alternative for the multi-component distributions used in phenomenological studies. We also generalize our framework for non-degenerate oscillators and encounter a Lerch-type generalization of the Barnes zeta function that, to the best of our knowledge, has not been studied in the context of nonadditive statistics and multiplicity studies before.
\end{abstract} 

\maketitle

\section{Introduction}

Many particles are produced in a high-energy collision event, and the total number of produced particles fluctuates from one event to another. The number of produced particles form a distribution $P(N)$ that (upon proper normalization) represents the probability that a given collision (event) produces exactly $N$ particles. 

%

For particles emitted from independent and identically distributed sources, the Poisson distribution provides a baseline for multiplicity fluctuations. However, It has been observed that the multiplicity distribution $P(N)$ obtained from $\mathrm{pp}$, $\mathrm{pA}$, and $\mathrm{AA}$ collisions are represented by distributions similar to the Negative Binomial Distribution (NBD) \cite{pnnbdprd} that is connected with significantly larger fluctuations in particle multiplicity than that showed by the Poisson distribution. 

In this article, we study the multiplicity distribution from the information theory approach \cite{infthnpb,infthjpg} using harmonic oscillators whose energies $\omega_i$ may correspond to the central values of momentum bins representing particle momenta. Hence, populating each bin amounts to changing the harmonic oscillator quantum numbers $n_i$, and the total number of particles produced in an event is given by $\sum_i n_i = N$.

With the help of the probability $P(N)$, we define an entropy that we maximize subject to a minimal (and physically relevant) set of constraints and find out the equilibrium set of probabilities $P(N)$. Using this approach for our oscillator model, we study the origin of the `Poissonian and beyond-Poissonian' distributions. First, we study a single-mode harmonic oscillator as the building block of our model, a perspective also utilized recently in Ref.~\cite{tsshoplb}. Eventually we generalize our results to multi-mode oscillators. The harmonic oscillator approach offers analytical tractability while capturing the essential physical scenario. 

Eqs.~\eqref{pnbgdeg}, \eqref{pntsdeg}, and \eqref{genbarzeta} highlight the main findings of the paper. 
\begin{itemize}
\item Starting from the Shannon entropy, we show that the equilibrium probability distribution $P(N)$ for a system of degenerate harmonic oscillators is represented by the NBD (Eq.~\ref{pnbgdeg}). Also, for infinitely many such oscillators, this distribution approaches the Poisson distribution when the mean multiplicity is kept fixed.
\item The NBD derived from the Boltzmann-Gibbs statistics assumes factorization of probabilities of micro-states, and we generalize this scenario by introducing the Tsallis entropy. We derive a generalized negative binomial distribution (gNBD) (Eq.~\ref{pntsdeg}) considering degenerate harmonic oscillators following Tsallis statistics. \textcolor{black}{We observe that, for the chosen datasets, the gNBD describes particle multiplicity data at high multiplicity region where a single NBD starts deviating from experimental data \cite{infthjpg}. This indicates that the gNBD may povide an alternative to the multi-component distributions used in the literature \cite{multcompnbd1,multcompnbd1a,multcompnbd2,multcompnbd3,multcompnbd4}. 
However, more comprehensive analysis is required to ascertain this conclusion.}

\item The generating function of the probability distribution is also a quantity of interest. While studying the generating function of the multi-mode non-degenerate oscillators within the Tsallis statistics framework, we encounter a special function that we call the `Lerch-type generalization' of the Barnes zeta function (Eq.~\ref{genbarzeta}). Some special cases of this function have been studied in physics \cite{ishiharazeta} and mathematics \cite{srivastavazeta}. However, to the best of our knowledge, the generalized Lerch Barnes zeta function reported in this article has not been considered earlier in the context of nonadditive statistics and multiplicity generating studies. It would be interesting to study the analytical structure of this function.
\end{itemize}

The rest of the paper is devoted to a detailed description of the model and their implications. In the following section, we consider the Boltzmann-Gibbs (BG) statistics and Shannon entropy. Later, we generalize the calculations by involving the Tsallis statistics (section \ref{Tsosc}) that introduces correlation among the oscillator modes.  \textcolor{black} {We detail phenomenological implications in section \ref{exp}, and we summarize and conclude in section \ref{summary}.}

\section{Boltzmann-Gibbs statistics, Oscillators, and Multiplicity Distributions}
\label{BGosc}

\subsection{ A single-mode oscillator}
Below we present the results that are already known by studying a single-mode oscillator in Boltzmann-Gibbs statistics.
By extremizing the Shannon entropy (see Appendix \ref{maxent}), subject to probability normalization and fixed mean energy constraints (expressed in terms of the energy of a micro-state $\epsilon$), the equilibrium set of probabilities is given by
\bea
P = \frac{\exp(-\beta \epsilon)}{\mathcal{Z}}.
\eea
For a single-mode harmonic oscillator, probability for $N$-th state with (normal ordered) energy $\epsilon =\omega N$ is
\begin{equation}
P(N)=\frac{1}{\mathcal{Z}}\exp(-\beta \omega N),
\label{PB}
\end{equation}
where the partition function is given by

\begin{equation}
\mathcal{Z}=\sum_{N=0}^{\infty}\exp(-\beta\omega N)=\frac{1}{1-\exp(-\beta\omega)}.
\label{ZB}
\end{equation}
Hence, the probability distribution in Eq.~\eqref{PB}

\begin{equation}
P(N)=\exp(-\beta\omega N)(1-\exp(-\beta\omega))
\label{PB1}
\end{equation}
is the geometric distribution with mean

\begin{equation}
\langle N \rangle=\sum_{N=0}^{\infty} N P(N)=\frac{1}{\exp(\beta\omega)-1}
\label{mB}
\end{equation}
and variance $ \text{Var}(N)= \langle N \rangle (1+ \langle N \rangle)$.

\subsection{Multi-mode oscillators, NBD, and Poisson}
\subsubsection{Probability distribution}
For a collection of $k$ harmonic oscillators of different frequencies $\omega_1,~\omega_2,.......\omega_k$ (with indices $n_1,~n_2,.......n_k$) in the Boltzmann-Gibbs statistics, the set of probabilities $P(\{n_i\})$ is given by
\bea
P(\{n_i\}) = \frac{e^{-\beta \sum_i \omega_i n_i}}{\mathcal{Z}_1\mathcal{Z}_2.......\mathcal{Z}_k},
\eea
where $\mathcal{Z}$ is the partition function. From the above equation, the multiplicity distribution $P(N)$ can be equated to
\bea
P \left(N=\sum_i n_i\right) = 
\sum_{\substack{n_1+n_2+\cdots+n_k=N\\
n_1,n_2,\ldots,n_k\geq0}}
e^{-\beta \omega_1 n_1} (1-e^{-\beta \omega_1})e^{-\beta \omega_2 n_2} (1-e^{-\beta \omega_2})......
\eea
Considering degenerate oscillators having $\omega_i \approx \omega$ and expressing $p\equiv \exp(-\beta \omega)$ 
, the distribution becomes as follows
\bea
P_{\text{D}}(N) &=& e^{-\beta \omega N}  (1-e^{-\beta \omega})^k \left( 
\sum_{\substack{n_1+n_2+\cdots+n_k=N\\
n_1,n_2,\ldots,n_k\geq0}}
1\right)
\nn\\
&=& \frac{(N+k-1)!}{N!(k-1)!} p^{N}  (1-p)^k.
\label{pnbgdeg}
\eea
Hence, we observe that our model involving degenerate oscillators in the BG statistics gives rise to the negative binomial distribution that describes multiplicity data obtained in many experiments.

\subsubsection{Generating function}
For the sake of completeness, we also study the generating function given below
\bea
G(z) = \sum_{N=0}^{\infty} P(N) z^N,
\eea
where $z$ is a dummy variable. In the BG oscillator model
\bea
G(z) 
&=& 
\frac{1}{\mathcal{Z}_1\mathcal{Z}_2.......\mathcal{Z}_k}  \left( \sum_{n_1=0}^{\infty} e^{-\beta n_1 \omega_1}z^{n_1} \right) .... \left( \sum_{n_k=0}^{\infty} e^{-\beta n_k \omega_k}z^{n_k} \right) \nn\\
&=& \prod_k \frac{1-e^{-\beta \omega_k}}{1-z e^{-\beta \omega_k}}.
\eea
For degenerate oscillators we obtain
\bea
G_{\text{D}}(z) &=& \left( \frac{1-e^{-\beta \omega}}{1-z e^{-\beta \omega}} \right)^k \nn\\
&=&
\left[1-\frac{p}{1-p} (z-1) \right]^{-k}.
\label{gzbgdeg}
\eea
The function $G(z)$ is the generating function for the NBD and is related to $P(N)$ by
\bea
P(N) = \left. \frac{1}{N!} \frac{d^N G(z)}{dz^N} \right|_{z=0}.
\eea
It is also worth mentioning that when $k \rightarrow \infty$, the NBD approaches the Poisson distribution when the mean multiplicity is kept fixed. 


\section{Nonadditive statistics, Oscillators, and Multiplicity Distributions}
In the BG statistics, we assume probability micro-states to be independent. However, for systems produced in high-energy collisions, fluctuations, long-range correlations, and memory effects may be prevalent, and the assumption of factorization may be relaxed. Hence, in this section we explore the statistical mechanics of oscillators in a generalized setting inspired by nonadditive statistics.
\label{Tsosc}
\subsection{A single-mode oscillator}

By extremizing the Tsallis entropy, subject to probability normalization and fixed mean energy constraints (expressed in terms of the energy of a micro-state $\epsilon$), the equilibrium set of probabilities is given by

\begin{equation}
P(N)=\frac{1}{\mathcal{Z}_q}\exp_q(-\beta \omega N)
\label{PT}
\end{equation}
where (for $q>0$, and $x>0$)
\begin{equation}
\exp_q(-x)=(1+(q-1)x)^{\frac{1}{1-q}}.
\label{eq}
\end{equation} 
The  normalization factor is given by the Hurwitz zeta function $\zeta$

\bea
\mathcal{Z}_q
&=&
\sum_{N=0}^{\infty} \left[ 1+(q-1)\beta\omega N \right]^{-\frac{1}{q-1}} 
\nn\\ 
&=&
\{(q-1)\beta\omega\}^{\frac{1}{1-q}}\zeta \left(\frac{1}{q-1}, \frac{1}{(q-1)\beta\omega}\right)
\label{ZT}
\eea
where  $\zeta(s,a)=\sum_{n=0}^{\infty}(n+a)^{-s}$. In stead of a geometric distribution (\ref{PB1}), now we obtain the Zipf-Mandelbrot distribution \cite{ZM1}
\begin{equation}
P(N)= \frac{1}{\mathcal{Z}_q }\left( 1+(q-1) \beta \omega N \right)^{\frac{1}{1-q}}.
\label{Mandelbrot}
\end{equation}
The mean multiplicity is given by

\begin{equation}
\langle N \rangle=\sum_{N=0}^{\infty} NP(N)=
\frac{\zeta\left( \frac{1}{q-1}-1,\frac{1}{(q-1)\beta\omega}\right)}{\zeta\left(\frac{1}{q-1}, \frac{1}{(q-1)\beta\omega} \right)}-\left( \frac{1}{(q-1)\beta\omega} \right)
\label{mT}
\end{equation}
For several more analytical results involving two and three oscillators and nonadditive statistics, please refer to Appendix.

\subsection{Multimode oscillators}

\subsubsection{Probability distribution}
Following Sec.~\eqref{BGosc}, we now study multiplicity distributions using multimode oscillators in nonadditive statistics. The set of nonadditive probabilities are given by
\bea
P(\{n_i\}) = \frac{1}{\mathcal{Z}_q} \left[ 1+ (q-1) \beta \sum_i n_i \omega_i \right]^{-\frac{1}{q-1}}
\eea
For a fixed $N= \sum_i n_i$, the distribution $P(N)$ is given by the following constrained summation
\bea
P(N) = \frac{1}{\mathcal{Z}_q} 
\sum_{\substack{n_1+n_2+\cdots+n_k=N\\
n_1,n_2,\ldots,n_k\geq0}}
\left[ 1+ (q-1) \beta n_1 \omega_1 + (q-1) \beta n_2 \omega_2 +.....  \right]^{-\frac{1}{q-1}}
\label{pnts}
\eea
The partition function in the denominator can be expressed in terms of the Barnes zeta function $\zeta_{\text{B}}$ \cite{ishiharazeta} as follows
\bea
\mathcal{Z}_q 
&=& 
\sum_{n_1,n_2,..=0}^{\infty} \left[ 1+ (q-1) \beta n_1 \omega_1 + (q-1) \beta n_2 \omega_2 + (q-1) \beta n_3 \omega_3+.... \right]^{-\frac{1}{q-1}} \nn\\
&=& 
\left[ (q-1)\beta \omega_{\text{ref}} \right]^{-\frac{1}{q-1}} 
\sum_{n_1,n_2,..=0}^{\infty}
\left( \lambda + a_1 n_1 + a_2 n_2 + a_3 n_3 +......\right)^{-\frac{1}{q-1}} \nn\\
&=&
\left[ (q-1)\beta \omega_{\text{ref}} \right]^{-\frac{1}{q-1}}
\zeta_{\text{B}} \left( \frac{1}{q-1},\lambda | \{a_i\} \right).
\label{zqtsmult}
\eea
In the above equation, $\omega_{\text{ref}}$ is a reference energy, $\{a_i\} \equiv (a_1,a_2,a_3,....) \equiv \{\omega_i/\omega_{\text{ref}}\}$, and $\lambda = \{(q-1)\beta \omega_{\text{ref}}\}^{-1}$.
If we assume degenerate oscillators such that $\omega_i \approx \omega~\forall~i$, and $\sum_i n_i=N$, Eq.~\eqref{pnts} becomes
\bea
P_{\text{D}}(N) 
&=& 
\frac{1}{\mathcal{Z}_q} 
\sum_{\substack{n_1+n_2+\cdots+n_k=N\\
n_1,n_2,\ldots,n_k\geq0}}
\left[ 1+ (q-1) \beta \omega N \right]^{-\frac{1}{q-1}} \nn\\
&=&
\frac{1}{\mathcal{Z}_q} \left[ 1+ (q-1) \beta \omega N \right]^{-\frac{1}{q-1}} \left(
\sum_{\substack{n_1+n_2+\cdots+n_k=N\\
n_1,n_2,\ldots,n_k\geq0}}
1 \right) \nn\\
&=&
\frac{1}{\mathcal{Z}_q} \left[ 1+ (q-1) \beta \omega N \right]^{-\frac{1}{q-1}} \frac{(N+k-1)!}{N!(k-1)!}
\label{pntsdeg}
\eea
The partition function $\mathcal{Z}_q$ can be evaluated as
\bea
\mathcal{Z}_q = \sum_{N=0}^{\infty} \left[ 1+ (q-1) \beta \omega N \right]^{-\frac{1}{q-1}} \frac{(N+k-1)!}{N!(k-1)!}.
\eea

\subsubsection{Generating function}
Since the generating function $\mathcal{G}(z)$ admits a sum of $N$ ranging from zero to infinity, we obtain the following
\bea
\mathcal{G}(z) 
&=& 
\sum_{N=0}^{\infty} P(N) z^N \nn\\
&=&
\frac{1}{\mathcal{Z}_q} \sum_{n_1=0}^{\infty} \sum_{n_2=0}^{\infty}.........\left[ 1+ (q-1) \beta n_1 \omega_1 + (q-1) \beta n_2 \omega_2 +.....  \right]^{-\frac{1}{q-1}} z^{n_1+n_2+....} \nn\\
&=&
\frac{1}{\mathcal{Z}_q}
[(q-1)\beta \omega_{\text{ref}}]^{-\frac{1}{q-1}}
\sum_{n_1=0}^{\infty} \sum_{n_2=0}^{\infty}.........
\frac{z^{n_1+n_2+....}}{(\lambda + a_1 n_1 + a_2 n_2 +.......)^{\frac{1}{q-1}}},
\label{gzts}
\eea
The summation in the above equation can be thought of as a Lerch-type generalization \cite{Bateman} of the Barnes zeta function that we can denote as $\Phi_{\text{B}}(z,1/(q-1),\lambda| \{a_i\})$. The generating function for non-degenerate oscillators is obtained as follows
\bea
\mathcal{G}(z) = \frac{\Phi_{\text{B}}(z,1/(q-1),\lambda| \{a_i\})}{\zeta_{\text{B}} (1/(q-1),\lambda| \{a_i\})}.
\label{genbarzeta}
\eea
To the best of our knowledge, combining a Barnes structure with a Lerch-like generalization has not been explicitly considered in the context of nonadditive statistics.
Some special cases have been considered in physics ($z=1$ case in Ref. \cite{ishiharazeta}) and mathematics ( $\{a_i\}= \{1\}$ case in Ref.~\cite{srivastavazeta}). 
It would be interesting to explore the analytical structure of such a generalization and other physical scenarios where it may arise.

For degenerate oscillators, the generating function in Eq.~\eqref{gzts} becomes
%
\bea
\mathcal{G}_{\text{D}}(z) 
&=& 
\frac{1}{\mathcal{Z}_q} \sum_{N=0}^{\infty} \frac{(N+k-1)!}{N!(k-1)!} \frac{z^N}{\left[ 1+ (q-1) \beta \omega N \right]^{\frac{1}{q-1}}}  \nn\\
&=&
\frac{1}{\mathcal{Z}_q (k-1)!} \sum_{N=0}^{\infty} (N+k-1)....(N+1) \frac{z^N}{\left[ 1+ (q-1) \beta \omega N \right]^{\frac{1}{q-1}}}  \nn\\
&=&
\frac{1}{\mathcal{Z}_q (k-1)!} \sum_{N=0}^{\infty} \left( a_{k-1} N^{k-1} + a_{k-2} N^{k-2}+.......+a_0\right) \frac{z^N}{\left[ 1+ (q-1) \beta \omega N \right]^{\frac{1}{q-1}}}  \nn\\
&=&
\frac{1}{\mathcal{Z}_q (k-1)!} \sum_{j=0}^{k-1} a_j \left( \sum_{N=0}^{\infty} \frac{N^j z^N}{\left[ 1+ (q-1) \beta \omega N \right]^{\frac{1}{q-1}}}  \right)
\label{gztsdeg}
\eea
The above formula can be expressed in terms of the Lerch's $\Phi$ function \cite{Bateman} (see Appendix \ref{lerch})
\bea
\mathcal{G}_{\text{D}}(z) =
\frac{[(q-1)\beta \omega]^{-\frac{1}{q-1}} }{\mathcal{Z}_q (k-1)!} \sum_{j=0}^{k-1} a_j \left(  
z\frac{\partial }{\partial z}
\right)^j 
\Phi \left(z,\frac{1}{q-1},\frac{1}{(q-1)\beta \omega}\right).
\eea
In the limit $q\rightarrow1$, the above equation converges to Eq.~\eqref{gzbgdeg}

\bea
\mathcal{G}_{\text{D}}(z) &\overset{q\rightarrow1}{=}&
\frac{(1-e^{-\beta \omega})^k}{(k-1)!} \sum_{j=0}^{k-1} a_j \left(  
z\frac{\partial }{\partial z}
\right)^j 
\left(
\frac{1}{1-z e^{-\beta \omega} }
\right)
\nn\\
&=&
\left( 
\frac{1-e^{-\beta \omega}}{1-z e^{-\beta \omega}}
\right)^k
\equiv
G_{\text{D}}(z)
\eea

\section{The Oscillator  Framework and Phenomenology}
\label{exp}

\begin{figure}[htbp]
    \centering
    \begin{minipage}{0.48\linewidth}
        \centering
        \includegraphics[width=\linewidth]{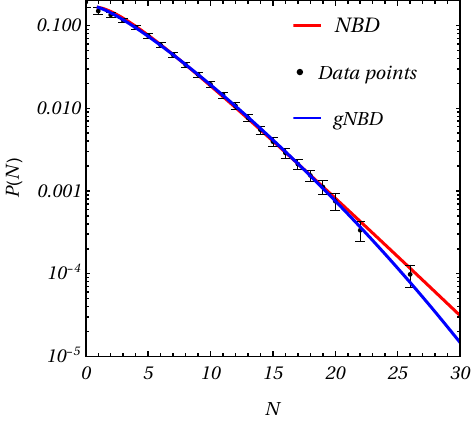}
        \caption{Multiplicity distribution measured by CMS \cite{CMS:2010qvf} in $\mathrm{pp}$ collisions at $\sqrt{s}=900$ GeV for $|\eta|<0.5$. NBD: $k$ = 1.5215, $\beta \omega$ = 0.34728. gNBD: $q$ = 0.95452,
        $k$ = 1.2834, $\beta \omega$ = 0.27701.}
        \label{cms1}
    \end{minipage}
    \hfill
    \begin{minipage}{0.48\linewidth}
        \centering
        \includegraphics[width=\linewidth]{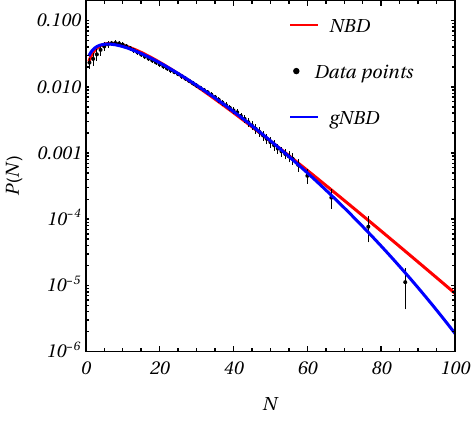}
        \caption{Multiplicity distribution measured by CMS \cite{CMS:2010qvf} in $\mathrm{pp}$ collisions at $\sqrt{s}=900$ GeV for $|\eta|<2$. 
        NBD: $k$ =1.8914, $\beta \omega$ = 0.11796. 
        gNBD: $q$ = 0.94548,
        $k$ = 1.6081, $\beta \omega$ = 0.089810.}
        \label{cms2}
    \end{minipage}
\end{figure}

\begin{figure}[htbp]
    \centering
    \begin{minipage}{0.48\linewidth}
        \centering
        \includegraphics[width=\linewidth]{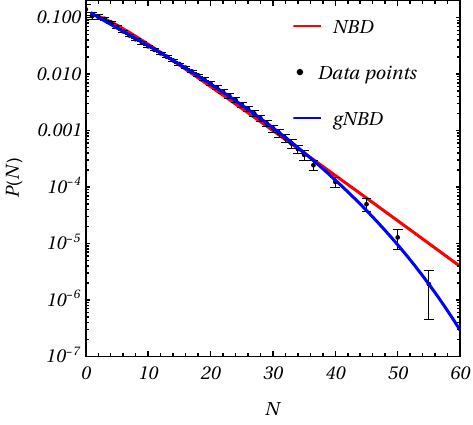}
        \caption{Multiplicity distribution measured by CMS \cite{CMS:2010qvf} in $\mathrm{pp}$ collisions at $\sqrt{s}=7000$ GeV for $|\eta|<0.5$.
        NBD: $k$ = 1.2784, 
        $\beta \omega$ = 0.19091. 
        gNBD: $q$ = 0.90797,
        $k$ = 0.94418, $\beta \omega$ = 0.12558.}
        \label{cms3}
    \end{minipage}
    \hfill
    \begin{minipage}{0.48\linewidth}
        \centering
        \includegraphics[width=\linewidth]{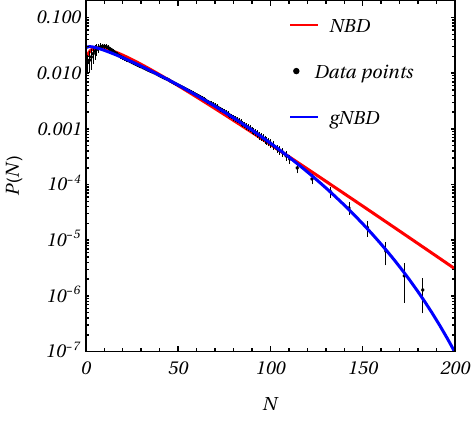}
        \caption{Multiplicity distribution measured by CMS \cite{CMS:2010qvf} in $\mathrm{pp}$ collisions at $\sqrt{s}=7000$ GeV for $|\eta|<2$.
         NBD: $k$ = 1.4116, 
        $\beta \omega$ = 0.054309. 
        gNBD: $q$ = 0.89235,
        $k$ = 1.0988, $\beta \omega$ = 0.035176.}
        \label{cms4}
    \end{minipage}
\end{figure}

\textcolor{black}{In this section, we compare the results obtained using the oscillator framework with the observed multiplicity distribution data. We find that the NBD obtained from the oscillator picture describes multiplicity distribution data up to certain multiplicity values (Fig.~\ref{cms1}), and starts deviating from the experimental observations for higher $N$ values. A similar observation has been reported in the literature. For example, Refs. \cite{multcompnbd1,multcompnbd1a} utilize a phenomenological two-component NBD for describing this deviation. However, the present work utilizes the NBD and a generalized NBD (Eqs.~\ref{pnbgdeg} and \ref{pntsdeg}) derived from an information theoretic model to describe experimental data. 
Comparison of Eqs.~\eqref{pnbgdeg} and \eqref{pntsdeg}, referred to as NBD and gNBD, with observed multiplicity distributions has been made in Figs.~\eqref{cms1}-\eqref{cms4}. The figures demonstrate that the while both NBD and gNBD describe low multiplicity region satisfactorily, they deviate at higher multiplicity region whereas the gNBD describes the data points. To demonstrate this deviation, we have chosen two rapidity windows, $|\eta|<0.5$ and $|\eta|<2$, at two different $\mathrm{pp}$ collision energies 900 GeV and 7 TeV. It is apparent that choosing a higher rapidity window helps explore higher multiplicity region where the effect of the generalized NBD derived in the manuscript is prominent. It is also noteworthy that comparison with experimental data leads to one of the fewer instances in high energy physics where the entropic parameter $q<1$.}

\section{Summary, conclusion, and outlook}
\label{summary}

In this work, we have proposed a harmonic oscillator-based model to describe multiplicity distributions in particle production processes. The probability distributions are derived from maximization of Boltzmann-Gibbs and generalized nonadditive entropies. \textcolor{black}{We observe that while both Boltzmann-Gibbs statistics and nonadditive statistics provide good description of chosen datasets at low multiplicity region, generalized NBD derived from nonadditive statistics describe data more efficiently in higher multiplicity region.}

The mathematical construction involving a system of degenerate oscillators provides a connection with the phenomenological NBD used in the literature. The phenomenological NBD is expressed in terms of the mean multiplicity $\langle N \rangle$ and the $k$ parameter 
. Since the total number $N$ is equal to $\sum_i n_i$ in the oscillator picture, the mean multiplicity $\langle N \rangle$ is equal to $\langle n_1 \rangle + \langle n_2 \rangle+ ....\langle n_k \rangle$. Each term $\langle n_i \rangle$ represents the Bose-Einstein single particle distribution (since we consider a bosonic oscillator) and is expressible in terms of the quantity $p = e^{-\beta \omega}$ (defined above Eq.~\ref{pnbgdeg}). For a system of $k$ degenerate oscillators, $\langle N \rangle = k p/(1-p)$, which implies that $\langle N \rangle/(\langle N \rangle +k) = p$. Hence, the  phenomenological $k$ parameter, that signifies dispersion, may correspond to the effective number of degenerate oscillators. Also, the factor  $\langle N \rangle/(\langle N \rangle +k) $ can be explained as the Boltzmann-Gibbs weight of adding one quantum to an oscillator, where $\langle N \rangle / k$ is the single particle distribution. 

\textcolor{black}{Analysis of experimental data using the oscillator framework reveals that the phenomenological parameter $q<1$ (which, when subtracted from 2, gives the entropic parameter $q$). A possible explanation for the sub-unity values of $q$ may be provided from the finite reservoir picture \cite{zwggwappb}. If we consider the observed system to be in contact with the unobserved part that may act as a reservoir, a nonadditive power law distribution                        may arise when the effective number of degrees of freedom of the bath is finite from the perspective of the system. For wider acceptance intervals (e.g. wider rapidity window), finite reservoir effects may be more prominent, potentially explaining the observations made in the paper. However, $q<1$ values introduce an upper cut-off in $N$ for real-valued probabilities, as we can see from Eq.~\eqref{pntsdeg}.}

\textcolor{black}{The oscillator picture has been utilized earlier to describe transverse momentum spectra \cite{tsshoplb}, and starting from the same framework, the present article describes multiplicity distributions.} This framework can also be extended to explore other observed quantities such as inelasticity, higher order multiplicity cumulants, and forward-backward correlation. We reserve these studies for the future.

\section{Appendix}
\subsection{Maximum entropy approach}
\label{maxent}

We maximize the Shannon entropy
\begin{equation}
S=-\sum_n P(n) \ln P(n)
\label{S}
\end{equation}
subject to some constraints.
Among all the discrete distributions with a specified mean value $\langle n \rangle$, the maximum entropy distribution is the geometric distribution.
Probability distribution $P(n)$ with mean $\langle n \rangle$ corresponds to maximum entropy when the quantity (expressed in terms of the multipliers $\lambda_1$ and $\lambda_2$)
\begin{equation}
F(P(n),\lambda_1,\lambda_2)=-\sum_nP(n)\ln P(n)+\lambda_1 \left(\sum_nP(n)-1 \right)
+\lambda_2 \left(\sum_n nP(n)- \langle n \rangle \right)
\label{FS}
\end{equation}
satisfies $\partial F/\partial P(n) =0$ for all $n$. The derivative yields $-1- \ln P(n)=\lambda_1+\lambda_2 n$ implying $P(n)=\exp(\lambda_1-1)\exp(\lambda_2 n)$.
Setting $r=\exp(\lambda_2)$, we obtain $P(n)=Cr^n$, where $C=\exp(\lambda_1-1)$. The normalization constraint $\sum_nP(n)=1$ implies $C=1-r$, and
\begin{equation}
P(n)=(1-r)r^n.
\label{PGD}
\end{equation}

Similarly, the Zipf-Mandelbrot distribution maximizes the Shannon entropy under the constraint
\cite{ZM2}
\begin{equation}
\langle \ln[1+(q-1)\beta\omega n] \rangle =\mu.
\label{ZMconst}
\end{equation}
Following the standard procedure, we consider the quantity
\begin{equation}
F(P(n),\lambda_1,\lambda_2)=-\sum_nP(n) \ln P(n)+\lambda_1 \left( \sum_nP(n)-1 \right)
+
\lambda_2 \left( \sum_nP(n)\ln [1+(q-1)\beta\omega n]-\mu \right).
\label{FSZM}
\end{equation}
From $\partial F / \partial P(n) =0$ for each $n$, and using the substitutions $\lambda_1-1 = \ln \lambda$ and $\lambda_2=1/(1-q)$, we obtain
\begin{equation}
P(n)=\lambda \left [1+(q-1) \beta \omega n \right ]^ {\frac{1}{1-q}}.
\label{PZM}
\end{equation}
The normalization constraint implies $\lambda^{-1}=Z= \left[ (q-1)\beta\omega \right]^{1/(1-q))}
\zeta \left( \frac{1}{q-1},\frac{1}{(q-1)\beta\omega}
\right)$.

For Tsallis entropy
\begin{equation}
H= - \sum_n \frac{P^q(n)-P(n)}{q-1},
\label{H}
\end{equation}
under constraint $\sum_n nP(n) = \langle n \rangle$,  we consider the quantity
\begin{equation}
F(P(n),\lambda_1,\lambda_2)=H+\lambda_1 \left(\sum_nP(n)-1 \right)
+
\lambda_2 \left(\sum_n nP(n)- \langle n \rangle \right).
\end{equation}
Setting $\partial F/\partial P(n)=0$ for each $n$ leads to
\begin{equation}
P(n)=q^{\frac{1}{1-q}}\left[ 1+(q-1)(\lambda_1 +\lambda_2 n)\right]^{\frac{1}{q-1}}.
\label{PTsallis}
\end{equation}
Substituting $q \rightarrow 2-q$ and reorganizing equation (\ref{PTsallis}) we have 

\begin{equation}
P(n)=\frac{1}{Z}\left[ 1+(q-1)\lambda n\right]^{\frac{1}{1-q}},
\label{PTsallis2}
\end{equation}
where $\lambda = \beta \omega$ for $\lambda_1=\frac{1-(2-q)Z^{q-1}}{q-1}$ and $ \lambda_2=\beta \omega (q-2) Z^{q-1}$.

\subsection{Two oscillator model: nonadditive statistics}
We consider two oscillators representing two bosonic particles and impose the constraint that the total number of particles is $N$. Hence
\bea
P(N) 
&=&
\frac{1}{\mathcal{Z}_q} \sum_{n_i; \sum n_i=N} \left[ 1+ (q-1) \beta n_1 \omega_1 + (q-1) \beta n_2 \omega_2 \right]^{-\frac{1}{q-1}} \nn\\
&=&
\frac{1}{\mathcal{Z}_q} \sum_{n_1=0}^N \left[ 1+ (q-1) \beta n_1 \omega_1 + (q-1) \beta (N-n_1) \omega_2 \right]^{-\frac{1}{q-1}} \nn\\
&=&
\frac{1}{\mathcal{Z}_q} \{(q-1)\beta (\omega_1-\omega_2)\}^{-\frac{1}{q-1}} 
\sum_{n_1=0}^N  \left[ \frac{1+(q-1)\beta N \omega_2}{(q-1)\beta (\omega_1-\omega_2)} +n_1 \right]^{-\frac{1}{q-1}} \nn\\
&=&
\frac{1}{\mathcal{Z}_q} \{(q-1)\beta (\omega_1-\omega_2)\}^{-\frac{1}{q-1}} 
\left[ \zeta \left( \frac{1}{q-1}, \frac{1+(q-1)\beta N \omega_2}{(q-1)\beta (\omega_1-\omega_2)} \right) 
- 
\zeta \left( \frac{1}{q-1}, 1+  \frac{1+(q-1)\beta N \omega_2}{(q-1)\beta (\omega_1-\omega_2)}   +N \right) \right]
\nn\\
\eea
where $\zeta$ represents the Hurwitz zeta function. The nonadditive partition function $\mathcal{Z}_q$ can be evaluated in terms of the two-parameter Barnes zeta function $\zeta_{\text{B}}^{(2)}$ as follows.
\bea
\mathcal{Z}_q 
&=& 
\sum_{n_1,n_2=0}^{\infty} \left[ 1+ (q-1) \beta n_1 \omega_1 + (q-1) \beta n_2 \omega_2 \right]^{-\frac{1}{q-1}} \nn\\
&=& 
\left[ (q-1)\beta \omega_{\text{ref}} \right]^{-\frac{1}{q-1}} 
\sum_{n_1,n_2=0}^{\infty}
\left( \lambda + a_1 n_1 + a_2 n_2\right)^{-\frac{1}{q-1}} \nn\\
&=&
\left[ (q-1)\beta \omega_{\text{ref}} \right]^{-\frac{1}{q-1}}
\zeta_{\text{B}}^{(2)} \left( \frac{1}{q-1},\lambda | \{a_i\} \right).
\label{zqts2}
\eea
In the above equation, $\omega_{\text{ref}}$ is a reference energy, $\{a_i\} \equiv (a_1,a_2) \equiv (\omega_1/\omega_{\text{ref}},~\omega_2/\omega_{\text{ref}})$, and $\lambda = \{(q-1)\beta \omega_{\text{ref}}\}^{-1}$.

\subsection{Three oscillator model: nonadditive statistics}
We present a three oscillator model comprising two bosonic and one fermionic oscillator. Hence $P(\{n_i\})$ can be written as:
\bea
P(\{n_i\}) = \frac{1}{\mathcal{Z}_q} \left[ 1+ (q-1) \beta n_1 \omega_1 + (q-1) \beta n_2 \omega_2 + (q-1) \beta n_3 \omega_3 \right]^{-\frac{1}{q-1}}.
\eea

If we consider the index $n_3$ to be associated with a fermionic oscillator it assumes only two values 0 and 1. Hence, if we impose the constraint that
$n_1+n_2+n_3 = N$, we have either $n_2=N-n_1$ (when $n_3$=0) or $n_2=N-n_1-1$ (when $n_3$=1). Hence, $P(N)$ can be written as a summation of two terms,
\bea
P(N) 
&=& 
\frac{1}{\mathcal{Z}_q^{(1)}} \sum_{n_1=0}^N \left[ 1+ (q-1) \beta n_1 \omega_1 + (q-1) \beta (N-n_1) \omega_2  \right]^{-\frac{1}{q-1}} \nn\\
&&+
\frac{1}{\mathcal{Z}_q^{(2)}} \sum_{n_1=0}^{N-1} \left[ 1+ (q-1) \beta n_1 \omega_1 + (q-1) \beta (N-n_1-1) \omega_2  + (q-1) \beta \omega_3 \right]^{-\frac{1}{q-1}} 
\nn\\
\eea

Following the previous section, the first sum can be evaluated as:
\bea
\frac{1}{\mathcal{Z}_q^{(1)}} \{(q-1)\beta (\omega_1-\omega_2)\}^{-\frac{1}{q-1}} 
\left[ \zeta \left( \frac{1}{q-1}, \frac{1+(q-1)\beta N \omega_2}{(q-1)\beta (\omega_1-\omega_2)} \right) 
- 
\zeta \left( \frac{1}{q-1}, 1+  \frac{1+(q-1)\beta N \omega_2}{(q-1)\beta (\omega_1-\omega_2)}   +N \right) \right].
\eea
The second sum can be evaluated as:
\bea
\frac{1}{\mathcal{Z}_q^{(2)}} \{(q-1)\beta (\omega_1-\omega_2)\}^{-\frac{1}{q-1}}
\times
\left[ \zeta \left( \frac{1}{q-1}, \frac{1+(q-1)\beta \{(N-1) \omega_2 +\omega_3 \}}{(q-1)\beta (\omega_1-\omega_2)} \right) \right. \nn\\
- 
\left. \zeta \left( \frac{1}{q-1}, \frac{1+(q-1)\beta \{(N-1) \omega_2 +\omega_3 \}}{(q-1)\beta (\omega_1-\omega_2)}   +N \right) \right].
\eea
$\mathcal{Z}_q^{(1)}$ and $\mathcal{Z}_q^{(2)}$ can be evaluated following the method outlined in the previous section.

\subsection{Generating functions: degenerate oscillators in nonadditive statistics}
\label{lerch}

Let us evaluate the quantity in the parentheses of Eq.~\eqref{gztsdeg}. When $j=0$, the quantity is proportional to the Lerch's $\Phi$ function
\bea
\sum_{N=0}^{\infty} 
\frac{z^N}{\left[ 1+ (q-1) \beta \omega N \right]^{\frac{1}{q-1}} }
= [(q-1)\beta \omega]^{-\frac{1}{q-1}} 
\Phi\left(z, \frac{1}{q-1}, \frac{1}{(q-1)\beta \omega} \right).
\eea
When $j=1$
\bea
\sum_{N=0}^{\infty} 
\frac{N z^N}{\left[ 1+ (q-1) \beta \omega N \right]^{\frac{1}{q-1}} }
= [(q-1)\beta \omega]^{-\frac{1}{q-1}} 
\left( z \frac{\partial }{\partial z} \right) \Phi.
\eea
When $j=2$
\bea
\sum_{N=0}^{\infty} 
\frac{N^2 z^N}{\left[ 1+ (q-1) \beta \omega N \right]^{\frac{1}{q-1}} }
&=&
\sum_{N=0}^{\infty} 
\frac{(N^2-N+N) z^N}{\left[ 1+ (q-1) \beta \omega N \right]^{\frac{1}{q-1}} }
\nn\\
&=&
\sum_{N=0}^{\infty} 
\frac{N(N-1) z^N}{\left[ 1+ (q-1) \beta \omega N \right]^{\frac{1}{q-1}} }
+
\sum_{N=0}^{\infty} 
\frac{N z^N}{\left[ 1+ (q-1) \beta \omega N \right]^{\frac{1}{q-1}} }
\nn\\
&=&
[(q-1)\beta \omega]^{-\frac{1}{q-1}} 
\left
[
\left( z^2 \frac{\partial^2 }{\partial z^2} \right) \Phi
+
\left( z \frac{\partial }{\partial z} \right) \Phi \right].
\nn\\
&=&
[(q-1)\beta \omega]^{-\frac{1}{q-1}} 
\left[
z \frac{\partial }{\partial z} \left( z \frac{\partial }{\partial z} \Phi \right) 
\right]
\nn\\
&=&
[(q-1)\beta \omega]^{-\frac{1}{q-1}} \left( z \frac{\partial }{\partial z} \right)^2 \Phi.
\eea
When $j=3$
\bea
\sum_{N=0}^{\infty} 
\frac{N^3 z^N}{\left[ 1+ (q-1) \beta \omega N \right]^{\frac{1}{q-1}} }
&=&
\sum_{N=0}^{\infty} 
\frac{(N^3-3N^2+2N+3N^2-2N) z^N}{\left[ 1+ (q-1) \beta \omega N \right]^{\frac{1}{q-1}} }
\nn\\
&=&
\sum_{N=0}^{\infty} 
\frac{(N-2)(N-1)N z^N}{\left[ 1+ (q-1) \beta \omega N \right]^{\frac{1}{q-1}} }
+
\sum_{N=0}^{\infty} 
\frac{3 N^2 z^N}{\left[ 1+ (q-1) \beta \omega N \right]^{\frac{1}{q-1}} }
-
\sum_{N=0}^{\infty} 
\frac{2N z^N}{\left[ 1+ (q-1) \beta \omega N \right]^{\frac{1}{q-1}} }
\nn\\
&=&
[(q-1)\beta \omega]^{-\frac{1}{q-1}} 
\left[
\left( z^3 \frac{\partial^3 }{\partial z^3} \right) \Phi
+
3 \left( z^2 \frac{\partial^2 }{\partial z^2} \right) \Phi
+
\left( z \frac{\partial }{\partial z} \right) \Phi 
\right] \nn\\
&=&
[(q-1)\beta \omega]^{-\frac{1}{q-1}} 
\left[
\left( z^3 \frac{\partial^3 }{\partial z^3} \right) \Phi
+
2 \left( z^2 \frac{\partial^2 }{\partial z^2} \right) \Phi 
+
\left( z^2 \frac{\partial^2 }{\partial z^2} \right) \Phi
+
\left( z \frac{\partial }{\partial z} \right) \Phi
\right]\nn\\
&=&
[(q-1)\beta \omega]^{-\frac{1}{q-1}} 
\left[
z\frac{\partial }{\partial z} \left( z^2 \frac{\partial^2 \Phi}{\partial z^2}  + z\frac{\partial \Phi}{\partial z}  \right) 
\right] \nn\\
&=&
[(q-1)\beta \omega]^{-\frac{1}{q-1}} 
\left[
z\frac{\partial }{\partial z}
\left\{
z\frac{\partial }{\partial z}
\left(
z\frac{\partial \Phi}{\partial z}
\right)
\right\}
\right]
\nn\\
&=&
[(q-1)\beta \omega]^{-\frac{1}{q-1}} 
\left(
z\frac{\partial }{\partial z}
\right)^3 \Phi.
\eea
By induction,
\bea
\sum_{N=0}^{\infty} 
\frac{N^j z^N}{\left[ 1+ (q-1) \beta \omega N \right]^{\frac{1}{q-1}} }
&=&
[(q-1)\beta \omega]^{-\frac{1}{q-1}} 
\left(
z\frac{\partial }{\partial z}
\right)^j \Phi.
\eea
Hence the generating function corresponding to the multiplicity distribution $P(N)$, considering that multiplicity $N$ is distributed among $k$ degenerate oscillators in a nonadditive thermal bath, is given by
\bea
{
\mathcal{G}_{\text{D}}(z) =
\frac{[(q-1)\beta \omega]^{-\frac{1}{q-1}} }{\mathcal{Z}_q (k-1)!} \sum_{j=0}^{k-1} a_j \left(  
z\frac{\partial }{\partial z}
\right)^j 
\Phi \left(z,\frac{1}{q-1},\frac{1}{(q-1)\beta \omega}\right)
.}
\eea

For the numerical purposes, expressing derivatives of the Lerch's $\Phi$ in terms of the shifted $\Phi$ functions will be more useful. Let us re-evaluate the parentheses of Eq.~\eqref{gztsdeg} for the $j=1$ case.
\bea
&&
\sum_{N=0}^{\infty} 
\frac{N z^N}{\left[ 1+ (q-1) \beta \omega N \right]^{\frac{1}{q-1}} }
=
\sum_{N=0}^{\infty} \frac{\left(N + \frac{1}{(q-1)\beta \omega} \right) z^N}{\left[ 1+ (q-1) \beta \omega N \right]^{\frac{1}{q-1}} }
-
\sum_{N=0}^{\infty} \frac{\frac{1}{(q-1)\beta \omega} z^N}{\left[ 1+ (q-1) \beta \omega N \right]^{\frac{1}{q-1}} }
\nn\\
&&=
[(q-1)\beta \omega]^{-\frac{1}{q-1}} \Phi \left( z, \frac{1}{q-1}-1, \frac{1}{(q-1)\beta \omega} \right)
-
[(q-1)\beta \omega]^{-\frac{1}{q-1}-1} \Phi \left( z, \frac{1}{q-1}, \frac{1}{(q-1)\beta \omega} \right).
\eea
Hence the generating function $\mathcal{G}_{\text{D}}^{(2)}(z)$ for the two oscillator ($k=2$) case is
\bea
\hspace{-10pt} 
\mathcal{G}_{\text{D}}^{(2)}(z)
&=&
\frac{1}{\mathcal{Z}_q} \left[ a_1 \sum_{N=0}^{\infty} 
\frac{N z^N}{\left[ 1+ (q-1) \beta \omega N \right]^{\frac{1}{q-1}} }
+
a_0 \sum_{N=0}^{\infty} 
\frac{z^N}{\left[ 1+ (q-1) \beta \omega N \right]^{\frac{1}{q-1}} }
\right]\nn\\
&=&
\frac{[(q-1)\beta \omega]^{-\frac{1}{q-1}} }{\mathcal{Z}_q}
\left[
	a_1 \Phi \left( z, \frac{1}{q-1}-1, \frac{1}{(q-1)\beta \omega} \right)
	-
	\frac{a_1 } {[(q-1)\beta \omega]} \Phi \left( z, \frac{1}{q-1}, \frac{1}{(q-1)\beta \omega} \right)
	\right. 
\nn\\
&& \left.+
a_0 
	\Phi \left( z, \frac{1}{q-1}, \frac{1}{(q-1)\beta \omega} \right)	
\right].
\eea 
Since for the $k=2$ case $g(N)=N+1 \Rightarrow a_0=a_1=1$
\bea
&&\mathcal{G}_{\text{D}}^{(2)}(z)
= 
\frac{[(q-1)\beta \omega]^{-\frac{1}{q-1}} }{\mathcal{Z}_q}
\left[
	\Phi \left( z, \frac{1}{q-1}-1, \frac{1}{(q-1)\beta \omega} \right)
	-
	\frac{1} {[(q-1)\beta \omega]} \Phi \left( z, \frac{1}{q-1}, \frac{1}{(q-1)\beta \omega} \right)
	\right. 
\nn\\
&& \left.+ 
	\Phi \left( z, \frac{1}{q-1}, \frac{1}{(q-1)\beta \omega} \right)	
\right].
\label{g2ana}
\eea
%
%
%
%

\begin{center}
    \textbf{\large{Acknowledgements}}
\end{center}
TB acknowledges funding from the European Union's Horizon Europe Programme, via the ERA Fellowship Grant Agreement number
101130816.

\end{document}